\documentstyle[12pt,psfig]{article}

\begin{document}
\vskip 1.5cm
\centerline{\large\bf Charmonium suppression from purely geometrical effects}
\medskip
\bigskip
\medskip
\centerline{\bf N. Hammon, L. Gerland, H. St\"ocker, W. Greiner\footnote{Work
supported by BMBF, DFG, GSI}}
\medskip
\bigskip
\centerline{Institut F\"ur Theoretische Physik,}
\centerline{Robert-Mayer Str. 10,}
\centerline{Johann Wolfgang Goethe-Universit\"at,}
\centerline{60054 Frankfurt am Main, Germany}
\medskip
\bigskip
\medskip
\bigskip
\centerline{Abstract}
\bigskip
{\small
The extend to which geometrical 
effects contribute to the production and suppression of the $J/\psi$ and $q\overline{q}$ 
minijet pairs in general is investigated
for high energy heavy ion collisions at SPS, RHIC and LHC energies. 
For the energy range under investigation, the geometrical effects 
referred to are shadowing and anti-shadowing, respectively.
Due to those effects, the parton distributions in nuclei deviate from the naive
extrapolation from the free nucleon result; $f_{A}\neq A f_{N}$.
The strength of the shadowing/anti-shadowing effect increases with the mass number.
Therefore it is interesting 
to see the difference between cross sections for e.g. $S+U$ vs. $Pb+Pb$ at SPS. 
The recent NA50 results for the survival probability of produced $J/\psi$'s has attracted 
great attention and are often interpreted as a signature of a quark gluon plasma.
This publication will present a fresh look on hard QCD effects
for the {\it charmonium production} level.
It is shown that the apparent suppression
of $J/\psi$'s must also be linked to the production process. 
Due to the uncertainty in the shadowing of gluons the suppression 
of charmonium states might not give reliable information on a created plasma phase at the 
collider energies soon available.\\
The consequences of shadowing effects for the $x_F$ 
distribution of $J/\psi$'s at 
$\sqrt s =20$ GeV, $\sqrt s =200$ GeV and $\sqrt s =6$ TeV are calculated for some relevant 
combinations of nuclei, as well as the $p_T$ distribution of minijets at midrapidity for 
$N_f=4$ in the final state.}
\newpage
\centerline{\bf 1. Introduction}

Since the advent of QCD in the 70's great emphasis was laid on the
existence of a phase transition of, yet unknown, order, being typical for
non-abelian gauge field theories. From lattice calculations it was emphasized that, 
at zero chemical
potential, a phase transition should show up at some temperature
$T_c \approx 150 - 200$ MeV when explicitly taking quarks into account. 
The value for $T_c$ is slightly higher for a pure gauge theory. 
Also, at non-zero chemical potential, as
suggested in the MIT bag model, one should access a phase transition due to
the increasing outward pressure of the partons inside the bag finally
leading to a deconfined phase. Due to the difficulties emerging when
considering dynamical fermions the work on non-zero chemical potential has
not yet reached the same level of success as that for $\mu =0$ in lattice QCD.\\
Now, in actual high energy heavy ion collisions the following scenario can occur.
Two streams of initially cold nuclear matter collide and may result in a plasma phase,
which is created within the transverse dimension of approximately the size of the 
overlapping nuclei. The plasma cools down to form hadronic degrees of freedom
in the subsequent expansion.
If one has this phase transition in mind one also has to confront the
question of its experimenatal detection.
Typical signatures under discussion are leptonic
(dilepton \cite{spieles} and photon \cite{nils} production due to the interactions among 
the quasi
free partons via the different QCD processes $q\overline{q}\rightarrow
\gamma g$, $g q \rightarrow \gamma q$, ...) and hadronic ones, such as the
suppression of $J/\psi$'s. Now, the QCD reactions in the plasma are not the
only source for leptons. One expects a large background coming from the
decay of $\pi^0$ and $\eta$ mesons. 
It therefore is necessary to
carefully handle this background by experimental methods such as invariant
mass analysis.\\
It is also obvious that the signatures have to give clear and powerful information on 
the plasma phase. Escpecially when looking at the hadronic signatures this does not
have to be the case as emphasized in \cite{hwa} where it was shown that gluon
depletion due to DGLAP splitting in the colliding nuclei can lead to the same results 
as the current experiments at NA50 \cite{NA50} show, which in turn implies that those 
experimental results propably have lost their meaning as a plasma signature at SPS.\\
\centerline{\bf 2. Production and suppression of the $J/\psi$}

\begin{figure}
\centerline{\psfig{figure=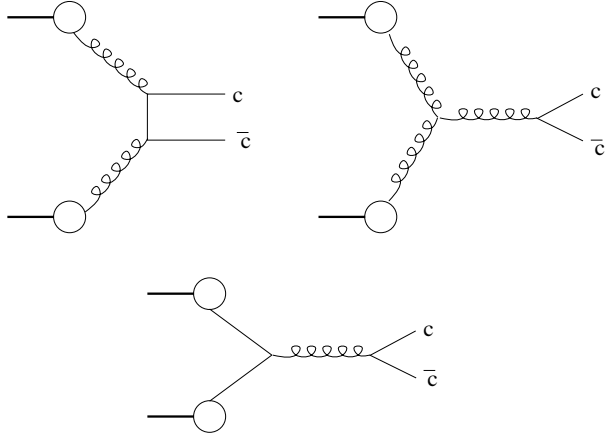,width=8cm}}
\caption{The various LO processes leading to the direct production of a
$c\bar c$ pair.}
\label{fig5-2}
\end{figure}

The $J/\psi$ is a $c\bar c$ bound state interacting via two forces in a
confined surrounding: a linear confining potential and a color-Coulomb 
interaction. In the plasma phase the linear potential is absent due to the high 
temperature leading to deconfinement.
Every color-charge is Debye-screened by a cloud of
surrounding quark-antiquark pairs which weaken the binding force between the
$c\bar c$ pair, thus reducing the color charge seen by the other (anti)quark. 
Since the density of the screening pairs rises strongly with
increasing temperature, the binding force gets weaker and weaker when the
temperature rises above $T_c$. As a result, the charm quark and antiquark drift away
from each other, so that finally no bound state formation is
possible in a plasma phase of high enough temperature \cite{matsui}.\\
However, the plasma phase is not the only source of suppression \cite{neubauer}. One
also has to take into account final state interactions for this hadronic
degree of freedom that are absent for leptonic signatures. Because the
$J/\psi$ is a verly weakly bound state, the interaction with nucleons and secondaries,
that are always present in a heavy ion collision, in addition significantly
lowers the survival propability for a $J/\psi$. It is obvious that such 
effects should increase with increasing mass number. One also expects the
phase transition to happen for the heavier nuclei. Therefore one has two effects both
increasing with the number of nucleons involved. This in turn implies that
the experiments have to be done with very high precision to disentangle 
those effects.\\
At this point another source of suppression comes into play that also
increases with the mass number and therefore has to be accounted for:
{\it nuclear shadowing}. This effect already enters on the
production level of the charmonium bound state. The former two effects,
namely suppression by melting in the plasma phase and comover activity,
enter only at a level when the $J/\psi$ already exists at later proper times $\tau$. 
Now the nuclear
shadowing effect appears when the charmonium is produced via the various
processes depicted in figure \ref{fig5-2}.\\
The total hidden charm
cross section in $pN$ collisions below the open charm threshhold is given by \cite{gavai} 
\begin{equation}
\sigma_{c\bar c}(s)=\int_{4m_{c}^{2}}^{4m_{D}^{2}} d\hat{s}
\int dx_A dx_B f_i (x_A) f_j (x_B) \hat{\sigma} (\hat s)\delta (\hat s - x_A x_B s)
\end{equation}
Here, $f_i$ and $f_j$ denote the parton densities and $\hat{\sigma}$ is the
cross section on the parton level, i.e. $q\bar q \rightarrow c \bar
c$, $gg\rightarrow c\bar c$.
%
%
The $c\bar c$ pair subsequently will turn into a color singlet by
interaction with the color field, induced in the scattering, the so-called
"color-evaporation" mechanism. In \cite{gavai} the $J/\psi$ production in a proton nucleon 
reaction was parametrized as 
\begin{equation}
\sigma_{pN\rightarrow J/\psi}(s)=f_{J/\psi}^{p}\sigma^{NLO}_{c\overline{c}}
\end{equation}
with $f_{J/\psi}^{p}=0.025$ from comparison with data \cite{schuler-cern}.
Here the production of the $J/\psi$ is described as proceeding via the NLO production of
a $c\bar c g$ state and a subsequent evaporation of the gluon.
(For a more detailed model also including the non-relativistic quarkonium model in the quarkonium- and
bottonium-nucleon cross section see \cite{lars}).\\
It is obvious that any changes in the parton densities will
result in changes of the $c\bar c$ production cross section. Because we know since the
EMC measurements \cite{EMC} that $f_i ^A \neq A f_i ^N$ this demands for some
further investigation. 
Here we will investigate the influence of the {\it nuclear gluon and quark distributions} 
on the $J/\psi$
production cross section by using a modified version of a parametrization based on a 
(impact parameter averaged)
data fit given in \cite{kari}. 
We will show the influence on the differential cross section
$d\sigma^{AB} /dx_F$ for $gg$ fusion and $q\bar q$ annihilation \cite{glueck} 
given by
\begin{eqnarray}
\frac{d\sigma^{AB}_{gg}}{dx_F}&=&\int_{4m_{c}^{2}}^{4m_{D}^{2}}
dQ^2 \frac{1}{Q^2}\frac{x_A x_B}{x_A + x_B} \hat{\sigma}^{gg\rightarrow c\overline c}
(Q^2)\nonumber\\
&\times & g^A(x_A,Q^2) g^B(x_B,Q^2)
\end{eqnarray}
\begin{eqnarray}
\frac{d\sigma^{AB}_{q\bar q}}{dx_F}&=&\sum_{q=u,d,s}\int_{4m_{c}^{2}}^{4m_{D}^{2}}
dQ^2 \frac{1}{Q^2}\frac{x_A x_B}{x_A + x_B} \hat{\sigma}^{q\bar q\rightarrow c\overline c}
(Q^2)\nonumber\\
&\times & \left[ q^A(x_A,Q^2) \bar q^B(x_B,Q^2) +q\leftrightarrow \bar q\right]
\end{eqnarray}
and on the minijet cross section
\begin{eqnarray}
\frac{d\sigma}{p_Tdp_Tdy_1dy_2}=2\pi x_A 
g^{A}(x_A,p_{T}^{2})x_B g^{B}(x_B,p_{T}^{2})\frac{d\hat{\sigma} ^{gg\rightarrow q\bar q}}
{d\hat{t}}
\end{eqnarray}
at midrapidity $y=y_1 =y_2 =0$.
We choose $m_c =1.5$ GeV and $m_D  =1.85$ GeV. The momentum fractions are given as
$x_{A,B} = 1/2 [\pm x_F + (x_{F}^{2}+4Q^2/s)^{1/2}]$ for the $x_F$ distribution 
and $x = 2p_T / \sqrt{s}$ for the minijets at midrapidity. 
We take all cross sections in leading order and do not include any K factor for higher
order contributions since we are mainly interested in relative effects. For the
parton distributions we choose the CTEQ4L parametrization.\\
The reason for our investigation is the following:
the recent NA50 data show a deviation from the tendency expected from earlier experiments
when the mass number of the involved nuclei is increased. 
Now, in \cite{hwa} it was shown that due to multiple scatterings between partons
the uncertainty in the survival propability gets so large that one cannot distinguish
whether the data found by NA50 is due to gluon splitting in the production phase or due 
to plasma absorption as claimed by several authors. Obviously, the originally good idea of
$J/\psi$ suppression as a good tool for plasma investigation seems to 
has lost its predictive power
at the available energies. It is therefore interesting to see what one can expect at
future colliders.

In the next part we will give some details of the parton
densities in nuclei in the energy regimes of SPS, RHIC, and LHC.\\ \\
\centerline{\bf 3. Nuclear shadowing and the connection to the $J/\psi$}

The history of the modification of nuclear structure functions, as compared to the free
nucleon ones,
is founded on the findings of the EMC group
that lead to the so-called EMC effect \cite{EMC} (even though one should say that
shadowing effects in principle have been known since the 70's \cite{caldwell}). 
This effect shows that $f_A \neq A f_N$, 
which implies that the parton density in the nucleus is not simply given by the nucleon
number times the respective parton density in the nucleon. Depending on the
frame (lab- or infinite momentum frame) one derives completely different interpretations 
for the 
nuclear structure functions and for the deviations from the naive $pp$-extrapolations. For 
typical values of the momentum transfer in a $pp$ reaction of $p_T=1-6$ GeV, where
perturbation theory should be applicable, one is in the so-called
anti-shadowing region for SPS and in the shadowing region for RHIC and LHC at midrapidity.
In the following we will shortly review the interpretation of shadowing in the two relevant frames
and will start with the lab frame description which is the {\it natural}
frame for typical deep inelastic scattering measurements off nuclei (at least from the
experimental setup point of view).\\

\begin{figure}
\centerline{\psfig{figure=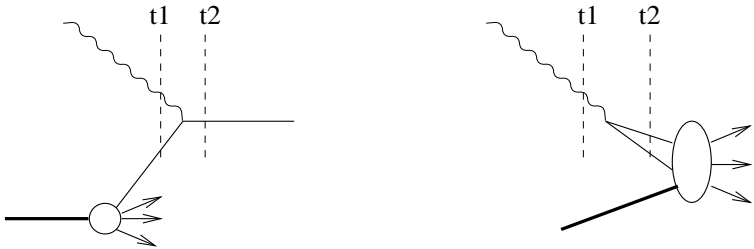,width=10cm}}
\caption{The two physical processes arising from the two possible time
orderings.}
\label{fig5-1}
\end {figure}

\centerline{\bf \small A. Lab frame description}
In the lab frame the expression {\it shadowing} immediately seems to imply a 
geometrical effect. When one speaks of something lying in the shadow of
another thing one means that the second body is not visible since the
first body is placed nearer, e.g. to some source of light. A similar reasoning can be
applied in the case when a lepton is scattered off a nucleus consisting of
many nucleons. The exchanged virtual photon does not (in the relevant $x$ range) interact 
individually with each nucleon but coherently with all nucleons or at least with a major
part of the nucleons inside the nucleus; some nucleons are therefore
lying {\it in the shadow} of other (surface) nucleons. As we will see later, this reasoning 
is linked to the momentum fraction of the struck parton inside the nucleon.
Because the momentum fraction is bound from above this 
interpretation is
limited to the explanation of shadowing and is not applicable for the reasoning of
anti-shadowing, the EMC effect or the Fermi-motion effect. Unfortunately
there is yet no single theory to understand the whole range of the momentum
fraction from $0\leq x_{Bj.} \leq 1$. For an excellent review of different models 
and interpretations see \cite{lonya}.\\
For a deep inelastic scattering process there exist two possible time
orderings for the interaction of a virtual photon with a nucleon or with a
nucleus: either the photon hits a quark inside the target (the so-called hand-bag graph)
or the photon creates a $q\bar q$ pair which then strongly interacts with the target.
Those two possible processes are depicted in figure \ref{fig5-1}.\\
As can be seen from the ratio of the amplitudes of the two processes one realizes that 
the diagram on the right hand side only contributes at small enough $x$ ($x\ll 0.1$).
At low $Q^2$ the interaction of the virtual photon with the nucleons inside
the nucleus happens via the low mass vector mesons $\rho$, $\omega$ and
$\phi$ as described in the vector meson dominance model (VMD) with the typical
spectral ansatz for the description of the fluctuation spectrum \cite{piller}. The
reduction in the quark density, manifesting itself in the shadowing ratio
$R_{F_{2}}=F_2^A/A\cdot F_2^N$, can then be understood in terms of a
multiple scattering series where the fluctuation interacts with more than one
nucleon over a coherence length of $l_c \approx 1/(2mx)$. 
At higher $Q^2$ the partonic degrees of freedom are probed; nevertheless,
shadowing is due to long distance effects and therefore always incorporates 
a strong non-perturbative component, even at large $Q^2$.
Also, the $q\overline{q}$ continuum
has to be taken in addition to the mesons giving rise to the generalized VMD model.\\
The interaction of the virtual photon with a nucleon can essentially be split up into 
two parts: the virtual photon with its quark-antiquark fluctuation and the interaction
of the fluctuation with the parton which happens via gluon exchange:
\begin{equation}
\sigma (\gamma^{*}N)=\int_{0}^1 dz \int d^2 r \left|
\psi(z,r)\right|^2\sigma_{q\bar q N}(r)
\end{equation}
where the Sudakov variable $z$ gives the momentum fraction carried by the quark (or antiquark).

The cross section for the
interaction of the fluctuation with the nucleon can be described in the DLA as
\begin{equation}
\sigma_{q\bar q N} = \frac{\pi^2}{3}r^{2}\alpha _s(Q'^2) x' g (x', Q'^2)
\label{eq1}
\end{equation}
where $x'=M_{q\bar q}^{2}/(2m\nu)$, $r$ is the transverse separation of
the pair and $Q'^2 =4/r^2$. 
Due to 
\begin{equation}
\left| \psi (z,r)\right| \sim \frac{1}{r^2}
\end{equation}
pairs with small transverse separation are favored. As can be seen from
(\ref{eq1}) this in turn implies a small cross section. This smallness is
compensated by the strong scaling violation of the gluon distribution in
the small $x$ region as $r$ ($Q'^2$) decreases (increases). 
In the Glauber eikonal approximation the interaction with the nucleus is
expressed in terms of the nuclear thickness funtion $T_A(b)$ as
\begin{equation}
\sigma_{q\bar q A}=\int d^2b\left( 1-e^{-\sigma_{q\bar qN} T_A(b)/2} \right)
\end{equation}
When there is a longitudinal momentum transfer appropriate to the production
of the hadronic fluctuation $h$ a phase shift behind the target results and the incident
wave ${\rm exp}(ik_z^{\gamma}z)$ is changed to $-\Gamma (b) {\rm exp}(ik_z^h z)$ with
$k_z^\gamma \neq k_z^h$ and the nucleus profile function $\Gamma (b)$. 
The phase shift $\Delta k_z = k_z^\gamma -k_z^h$
in turn gives rise to the coherence length $l_c \approx 1/\Delta k_z$. 
When now $l_c \gg 2R_A$ (i.e.~for small momentum transfers $\Delta k_z$), the hadronic 
fluctuation interacts coherently with all
nucleons inside the nucleus, Glauber theory is valid and a reduction in the
cross section results (for further details we refer to \cite{lonya,piller,ina,glr}). 
\begin{figure}
\centerline{\psfig{figure=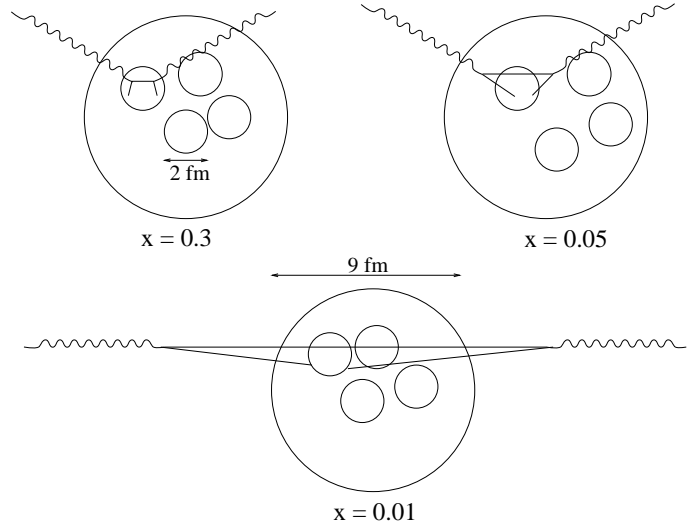,width=9cm}}
\caption{Graphic representation of the lab frame interpretation of shadowing.}
\label{fig5-3}
\end{figure}  
For an illustration 
of the effect see figure \ref{fig5-3}.\\
\centerline{\bf \small B. Infinite momentum frame description}\\
In the infinite momentum frame a completely different mechanism is employed. The key
idea here is the fusion of partons giving rise to a process that competes with parton 
splitting expressed in the DGLAP equations.
This idea was first formulated in \cite{glr} and later
proven in \cite{mq}. In the following we will give the main ideas and
conclusions of the parton fusion model \cite{close}.\\
As is known, in the infinite momentum frame the Bjorken variable
$x_{Bj}$ is interpreted as the momentum fraction of a parton with respect
to the mother nucleon. When now, inside a nucleus, the longitudinal wavelength of a 
parton exceeds the Lorentz-contracted size of a nucleon or the inter-nucleon
distance $2R_N$, then partons originating from different nucleons can "leak out" and fuse. This
effect can be estimated from $1/(xP) \approx 2R_N M_N/P$ to show up at
$x$ values smaller than $x\approx 0.1$. As a result of the parton-parton fusion partons are
"taken away" at smaller values of $x$ and "shoveled" to larger values of  
$x$ where anti-shadowing appears to guarantee momentum conservation.
As a result of the parton fusion the $x$-range for the
measured structure function is expanded to values $x>1$. Hereby, 
an alternative description of Fermi-motion is achieved.
In the lab frame interpretation the saturation of shadowing was interpreted
in terms of a coherence length larger than the nucleus. Here, the saturation
towards smaller $x$ values is interpreted in terms of the longitudinal
parton wave length exceeding the size of the nucleus. In that sense the
infinite momentum frame interpretation of shadowing is formulated in terms of
variables that are inherent to the nucleus and there is no need for a
scattered lepton or a collision.\\
In addition to the longitudinal shadowing one expects an additional
shadowing effect from the transverse fusion of partons: for sufficiantly
small values of $x$ and/or $Q^2$ the total transverse occupied area of the partons
becomes larger than the transverse area of the nucleon. This happens (e.g.~for gluons) 
when $xg(x)\geq Q^2 R^2$ where the transverse size of a parton
is $1/Q^2$ and $R$ is the nucleon radius. The depletion in the gluon and sea-quark 
densities arising from that process are expected at values $x\leq 0.01$.\\
\centerline{\bf 4. The used parametrization}

\begin{figure}
\centerline{\psfig{figure=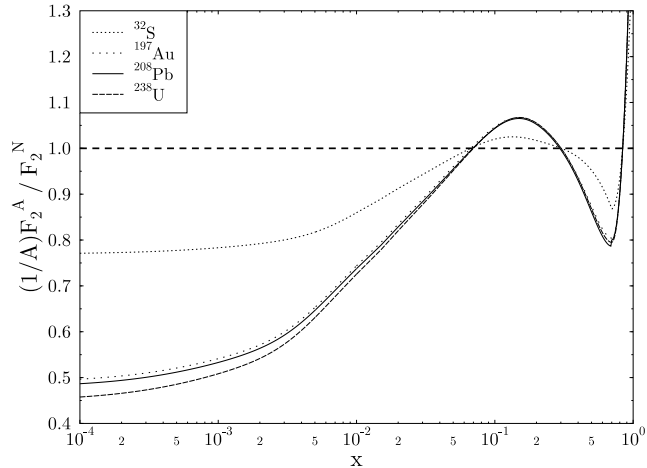,width=12cm}}
\caption{Fit to the data for various nuclei at $Q_0 =2$ GeV as given by Eskola.}
\label{fig5-4}
\end{figure}

In \cite{kari} a fit to the E772 \cite{E772}, NMC \cite{NMC} and SLAC \cite{SLAC} data 
was given as a parametrization for the ratio $R_{F_2}=F_2^A/A\cdot F_2 ^N$ (see figure
\ref{fig5-4}).
\begin{figure}
\centerline{\psfig{figure=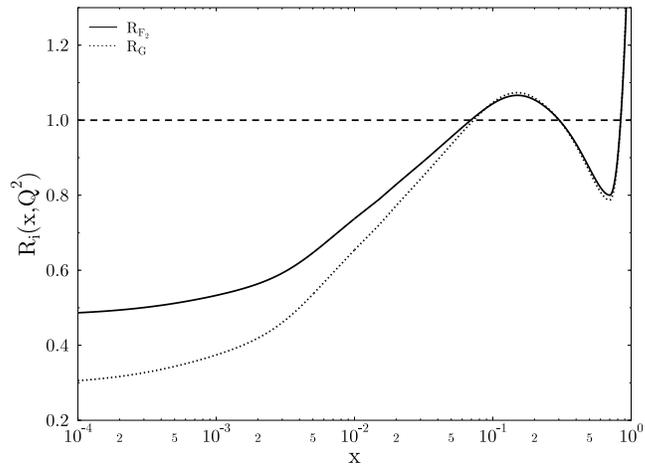,width=12cm}}
\caption{Initial gluon and quark shadowing parametrization at 
$Q^2 = 4$ GeV$^2$ for $^{197} Au$ and $^{208} Pb$.}
\label{fig5-4gluon}
\end{figure}  
\begin{figure}
\centerline{\psfig{figure=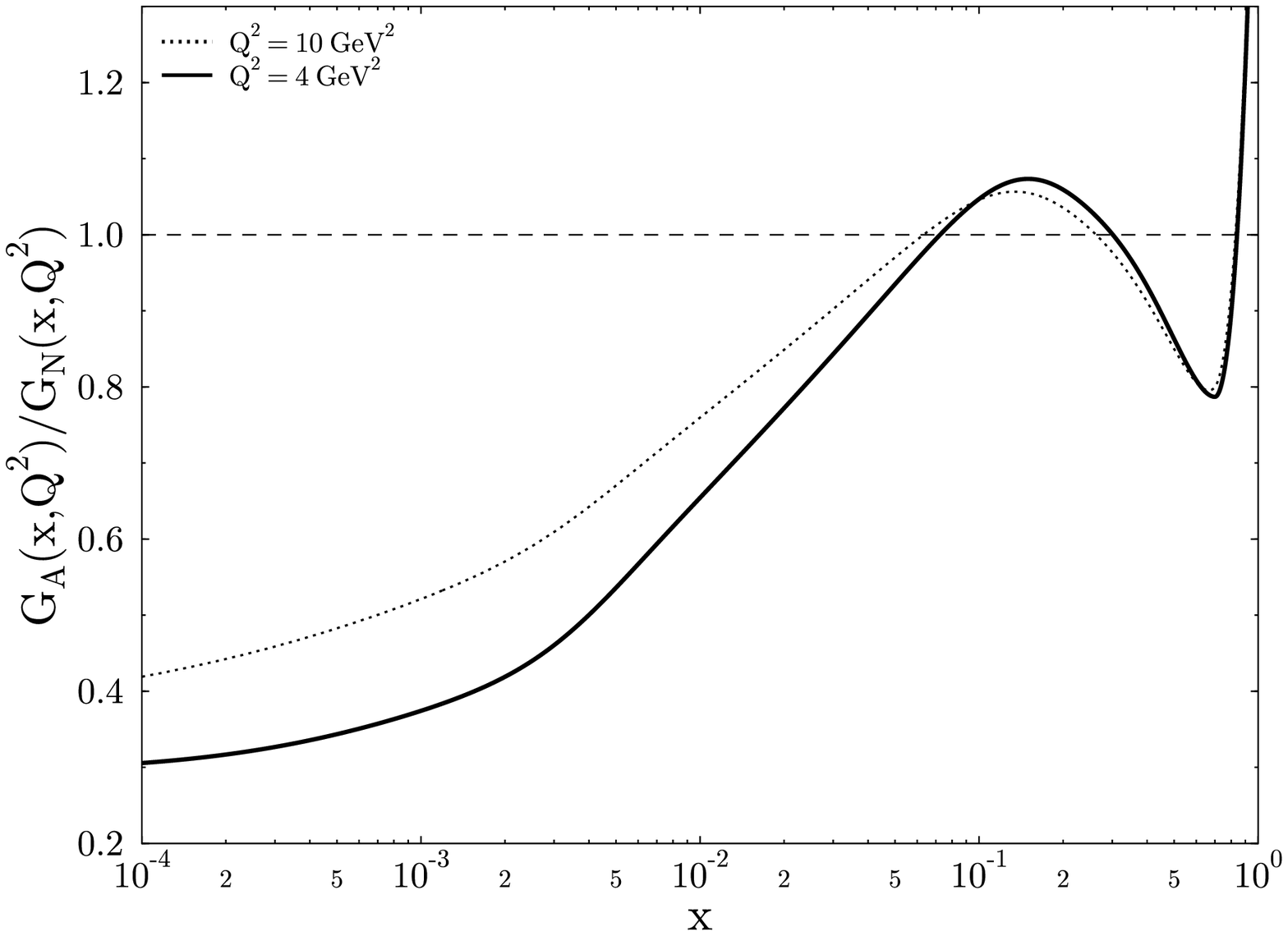,width=12cm}}
\caption{Gluon shadowing ratio evolved to $Q^2 = 10$ GeV$^2$ 
with DGLAP without fusion terms. Due to the narrow range 
$4m_{c}^{2}\leq Q^2 \leq 4m_{D}^{2}$ in the interpretation
we use the ratio at some fixed intermediate scale $Q^2 =10$ GeV$^2$.}
\label{fig5-4evolvgluon}
\end{figure}
Now this parametrized ratio cannot simply be multiplied with all the individual 
parton distributions entering the formulas. One has to make a distinction between the 
valence and the sea quarks and also needs a different ratio for the gluons.
Our results for $d\sigma^{AB} /dx_F$ are based on
the shape of the ratio given in \cite{kari} at the initial momentum transfer $Q_0=2$ GeV. 
Up to now, the production processes were often calculated by using the measured shadowing
ratio $R_{F_2}$. From the lab frame interpretation we know that the cross section for the
interaction of a gluon pair is larger than the one for the interaction of the quark-antiquark
pair ($\sigma^{pert.}_{ggN}=9/4 \sigma^{pert.}_{q\bar qN}$). The same tendency can be found in the parton 
fusion model. In \cite{kumano} calculations in the parton fusion model for $^{118} Sn$ 
showed an impact parameter averaged gluon shadowing that 
is as twice as strong ($R_G \approx 0.34$) as the sea quark shadowing at $x=10^{-3}$ 
and $Q^2 =5$ GeV$^2$ already for this light nucleus. To account for the much 
stronger gluon shadowing we therefore modified the parametrization given in \cite{kari}.\\

\begin{figure}
\centerline{\psfig{figure=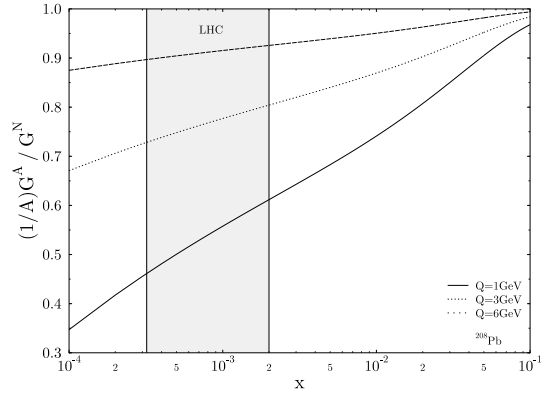,width=10cm}}
\centerline{\psfig{figure=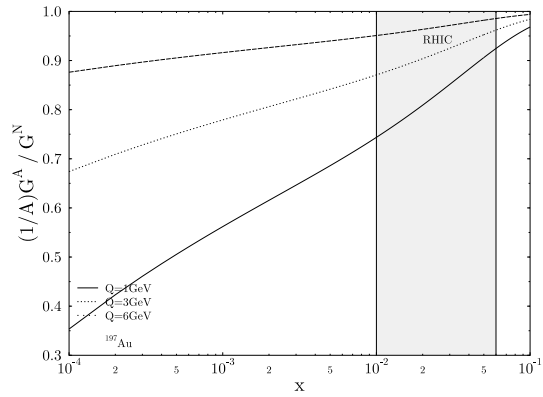,width=10cm}}
\caption{Gluon ratios corresponding to the various energy and transverse momentum regimes.}
\label{fig5-78}
\end{figure}

\begin{figure}
\centerline{\psfig{figure=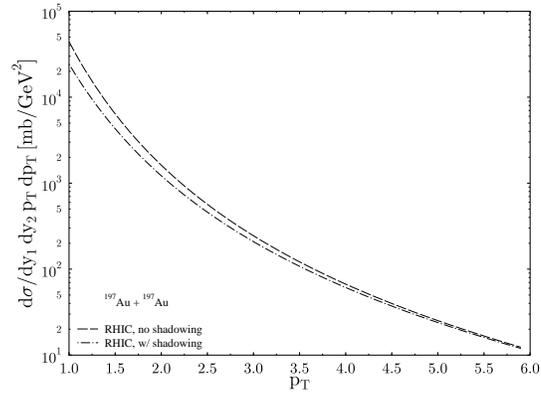,width=10cm}}
\centerline{\psfig{figure=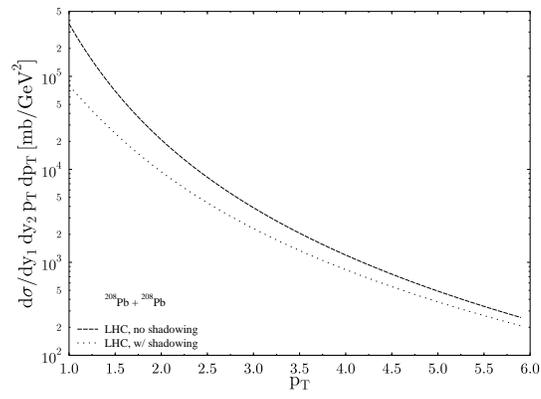,width=10cm}}
\caption{Minijet cross sections for creation of \protect$q\overline{q}$ pairs 
for RHIC and LHC.}
\label{fig5-1112}
\end{figure}

In the lab frame, the relevant range for the coherence length to produce the shadowing 
effect is $l_0 = r_{NN}\approx 1.8~fm \leq l_c = 1/(2mx) \leq 2 R_A$.
For $l_c \gg 2R_A$, corresponding to $x\ll 0.1~fm /1.1~fm A^{1/3}$, the shadowing of
gluons
at some initial scale at fixed impact parameter behaves as \cite{lonya1,mark}
\begin{equation}
\frac{A_{eff}}{A} = \frac{2-2({\rm exp} -R/2)}{R}
\end{equation}
where $R=T(b)\cdot \sigma_{eff}$. For the interaction of the $q\bar q$ pair one finds
$\sigma_{eff,q\bar q} \approx 14~mb$. which approximately corresponds to the $\rho N$
cross section. We here assume that the perturbative factor $9/4$ is valid also for the
non-perturbative regime and therefore choose $\sigma_{eff,gg} \approx 30~mb$. At $b=0$
and for $Pb$ one therefore has a maximum amount of shadowing of $A_{eff}/A \approx 0.39$
which is approximately $15\%$ smaller than the $b$-averaged result.
Because the two different scenarios (lab- or infinite momentum frame) give such different 
results, $R_G\approx 0.39$ \cite{lonya1,mark} vs. $R_G\ll 0.3$ for heavy nuclei at small
$x$, we decided to choose some intermediate value as a starting point for the DGLAP
evolution.\\
We therefore employ the curves shown in figure
\ref{fig5-4gluon} to account for the large difference in the quark- and gluon shadowing
ratios. Due to the large uncertainty of the initial $R_G$ we choose the same
ratio for $Au$ and $Pb$ at $Q^2 =10$ GeV$^2$ as shown in figure \ref{fig5-4evolvgluon}.
Also, we again want to emphasize that the commonly used shadowing ratios, 
only account for impact parameter averaged measurements in DIS reactions.
Therefore our results should be seen as for central events only because the production 
mechanism in very peripheral collisions should produce significantly smaller rates
with significantly smaller influences from shadowing effects \cite{emelyanov}.\\
For the minijet cross section we used a $Q^2$-dependent parametrization given in 
\cite{ina} to account for the larger $p_T$ region.\\
\centerline{\bf 4. Results}
\\
We will first present the results for the minijet cross section including 
only processes $i,j\rightarrow k,l$ with $i,j = g$ and $k,l = q\overline q$ with
four flavors in the final channel (due to the dominance of the $gg$ fusion process
annihilation processes are neglected at RHIC and LHC at midrapidity). 
For the $x_F$ distribution we used our modified
version of the parametrization in \cite{kari} but for the minijets we
used an impact parameter dependent parametrization with {\bf b} =0 \cite{ina} 
shown in figure \ref{fig5-78}. This parametrization is applicable here since we are in 
the pure shadowing region where the generalized VMD approach used to derive it is
applicable (even though one should say that the Glauber ansatz should only 
be valid up to values $x\sim 10^{-2}$ as restricted by the eikonal approximation).\\
The regions of the momentum fractions corresponding to the momentum range 
$1$ GeV $<p_T<$ $6$ GeV
for RHIC ($\sqrt s =200$ GeV) and LHC ($\sqrt s =6$ TeV) are represented in 
figure \ref{fig5-78} as shaded areas.
The results for the cross sections for RHIC and LHC are given in figure \ref{fig5-1112}.
In that calculation all quark antiquark pairs ($k,l=q\overline q$) up to the bottom
threshhold were taken into account, i.e. $N_f=4$ in the final state. One clearly sees the deviation
having its origin in the shadowing of the nuclear parton distribution. As expected, the 
shadowing effect decreases as $p_T$ increases due to the momentum fraction
$x=2p_T/\sqrt{s}$.\\

Next we will present the results for the $d\sigma^{AB}/dx_F$ cross sections.
We first calculated the proton-proton cross sections to show the dominance
of the $gg$ fusion process over the $q\overline q$ annihilation process 
at small $x_F$ (see figure \ref{fig5-151617}):
\begin{figure}
\centerline{\psfig{figure=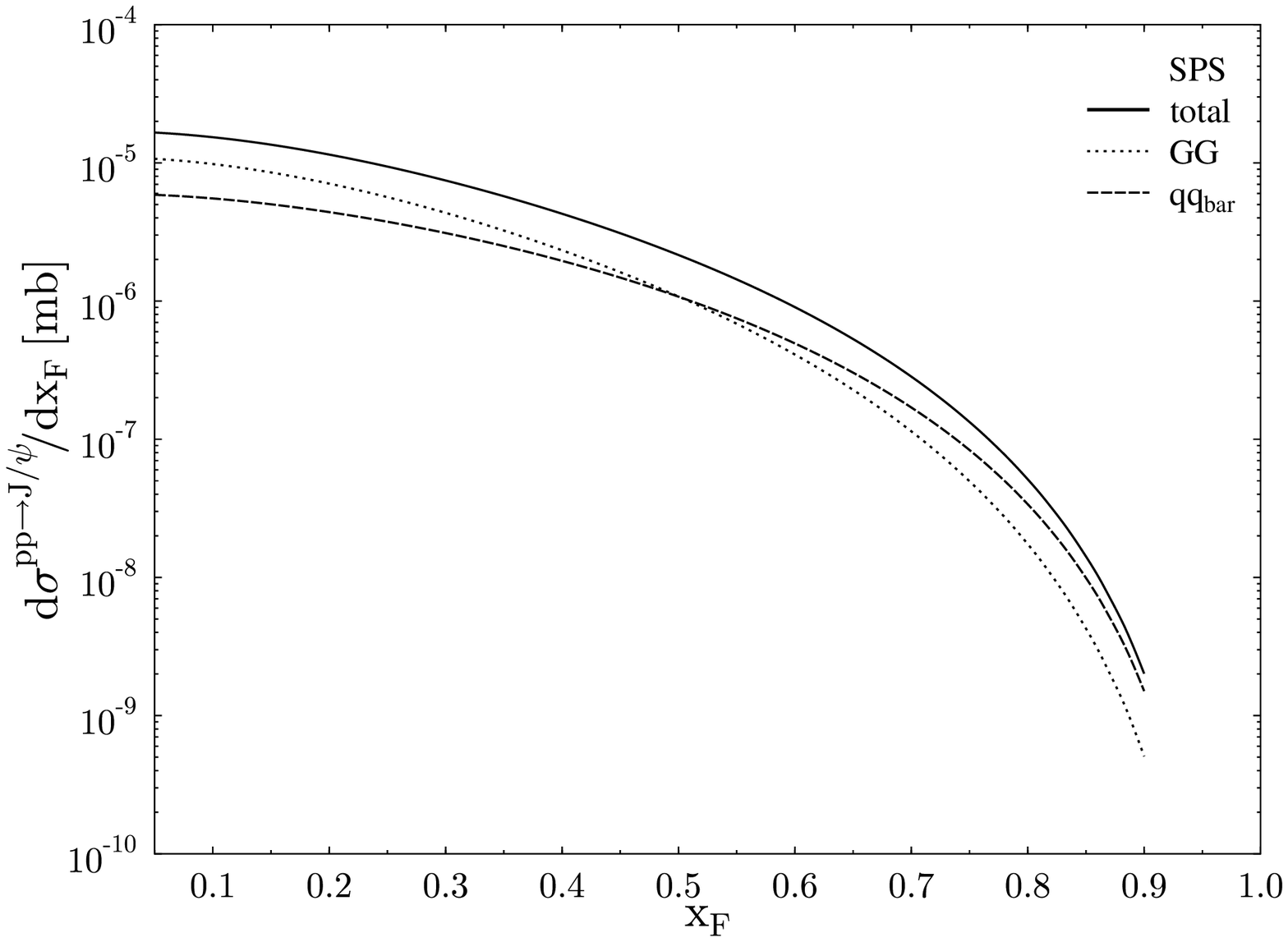,width=10cm}}
\centerline{\psfig{figure=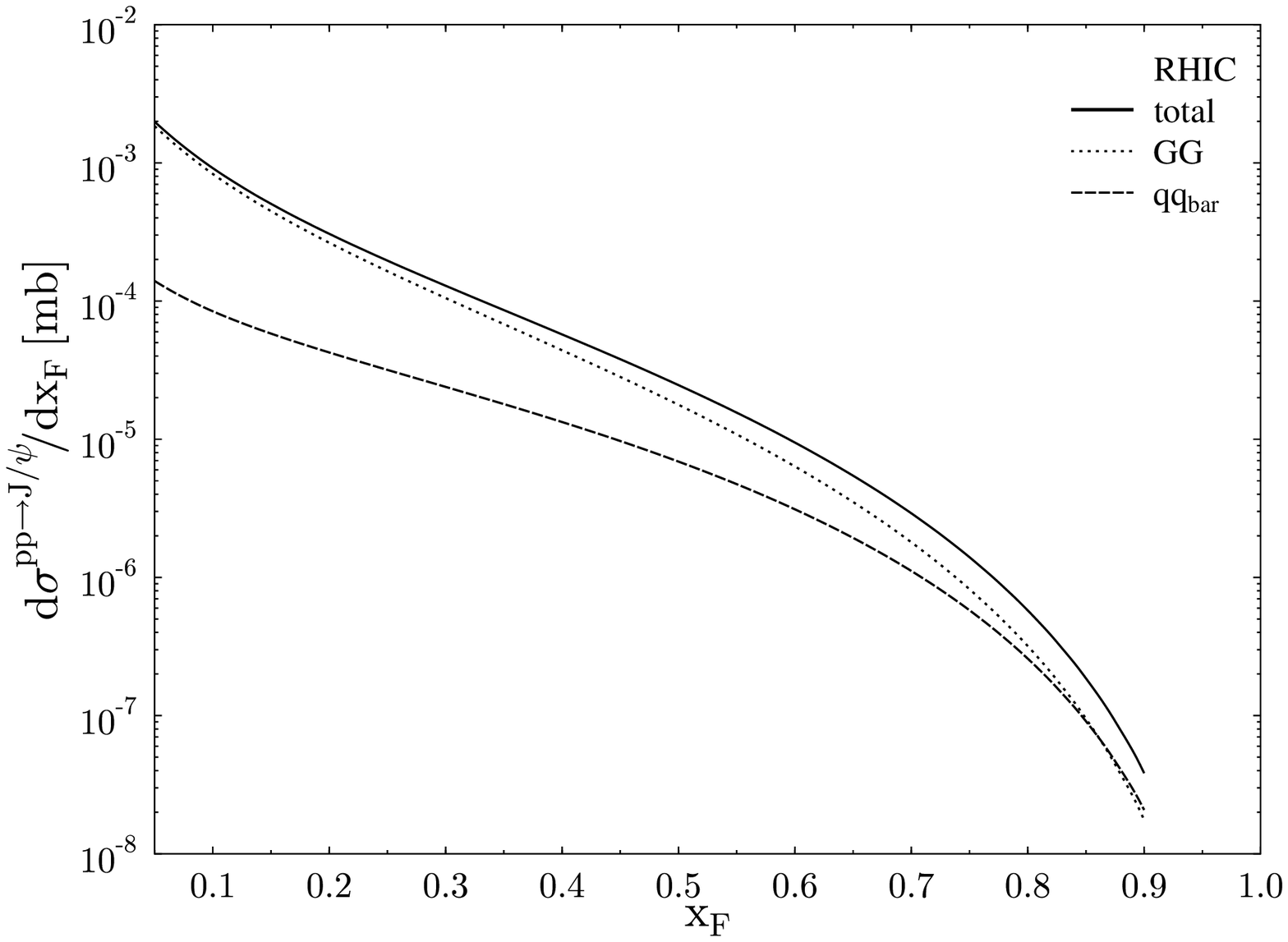,width=10cm}\psfig{figure=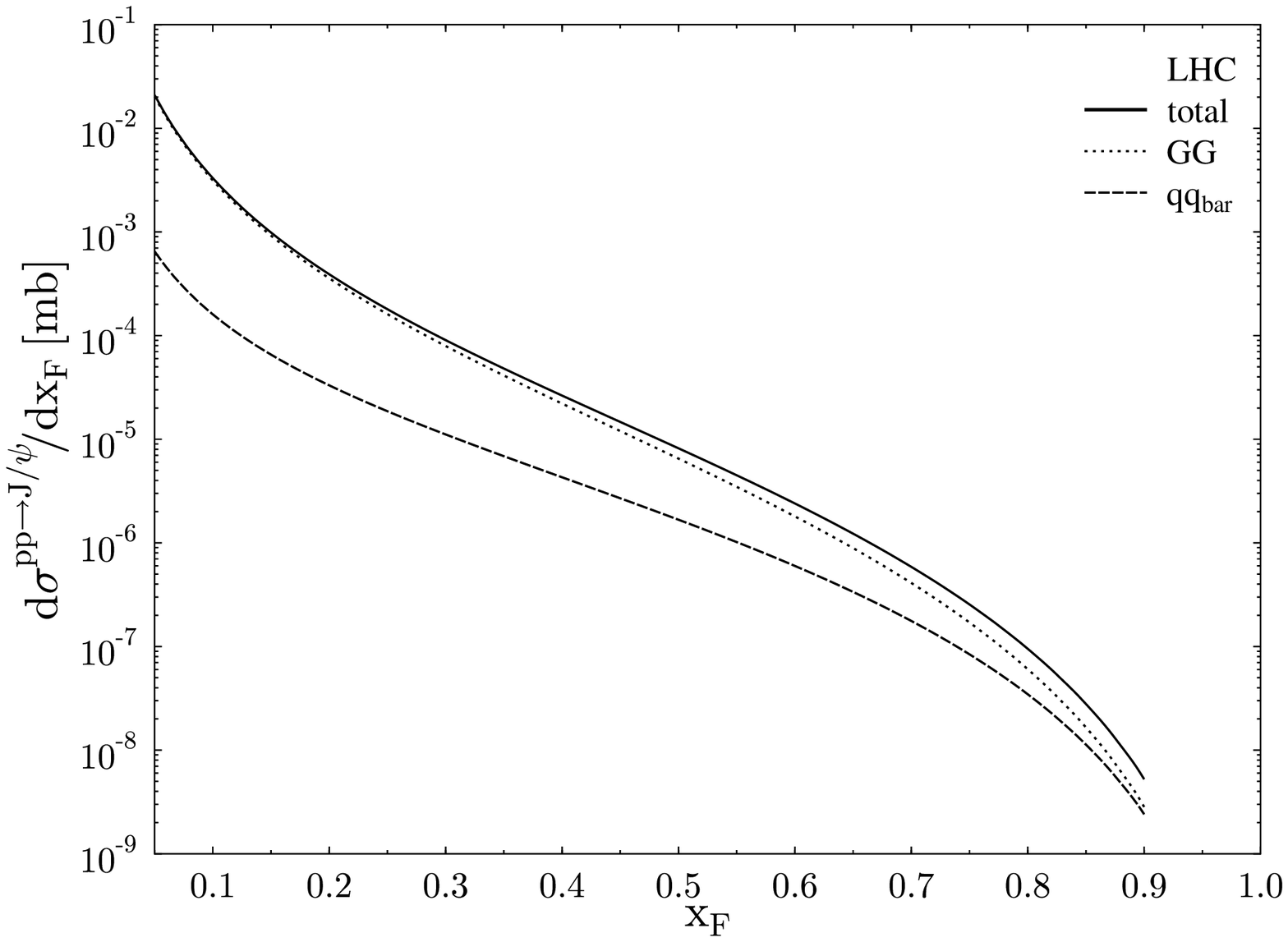,width=10cm}}
\caption{$d\sigma^{pp\rightarrow J/\psi}/dx_F$ for SPS, RHIC and LHC.}
\label{fig5-151617}
\end{figure}
The results for the cross sections for $S+U$ and $Pb+Pb$ at SPS at
$\sqrt s =20$ GeV, $Au+Au$ at RHIC at $\sqrt s =200$ GeV, and $Pb+Pb$ at LHC
at $\sqrt s =6$ TeV are presented in figure \ref{fig5-181920}. In this case we were restricted to our modified version of the 
parametrization of \cite{kari} due to the integration reaching up to momentum fractions
$x>0.1$, not allowing us to use the same parametrization as for the minijet production.
At SPS energies one clearly sees the different regions of the parametrization 
entering the cross section. At small $x_F$ one has the enhancement due to the
antishadowing which is followed by the depletion due to the EMC region at larger $x$
and finally one can identify the Fermi motion effect as $x_F\rightarrow 1$.
The effects are clearly stronger for $Pb+Pb$ than for $S+U$ (compare 
figure \ref{fig5-4}). To get an impression of the relative strength of the nuclear
modifications in the respective nuclei we calculated the ratios of the shadowed to
unshadowed cross sections in figure \ref{fig5-181920ratio}. 
The difference between $Pb+Pb$ and $S+U$ at SPS energies in principle is only small;
in the relevant region of small $x_F$, where the cross section has not dropped yet too much,
the charmonium production in $Pb+Pb$ is slightly larger than in $S+U$ ($\approx 8\%$).
For RHIC one is in the shadowing region. The suppression strongly varies over the $x_F$
range between $\approx 0.6-0.35$. At LHC an even stronger suppression is found due to
the smaller momentum fractions entering the shadowing ratios. Here the suppression is 
$\approx0.3-0.5$.
\begin{figure}
\centerline{\psfig{figure=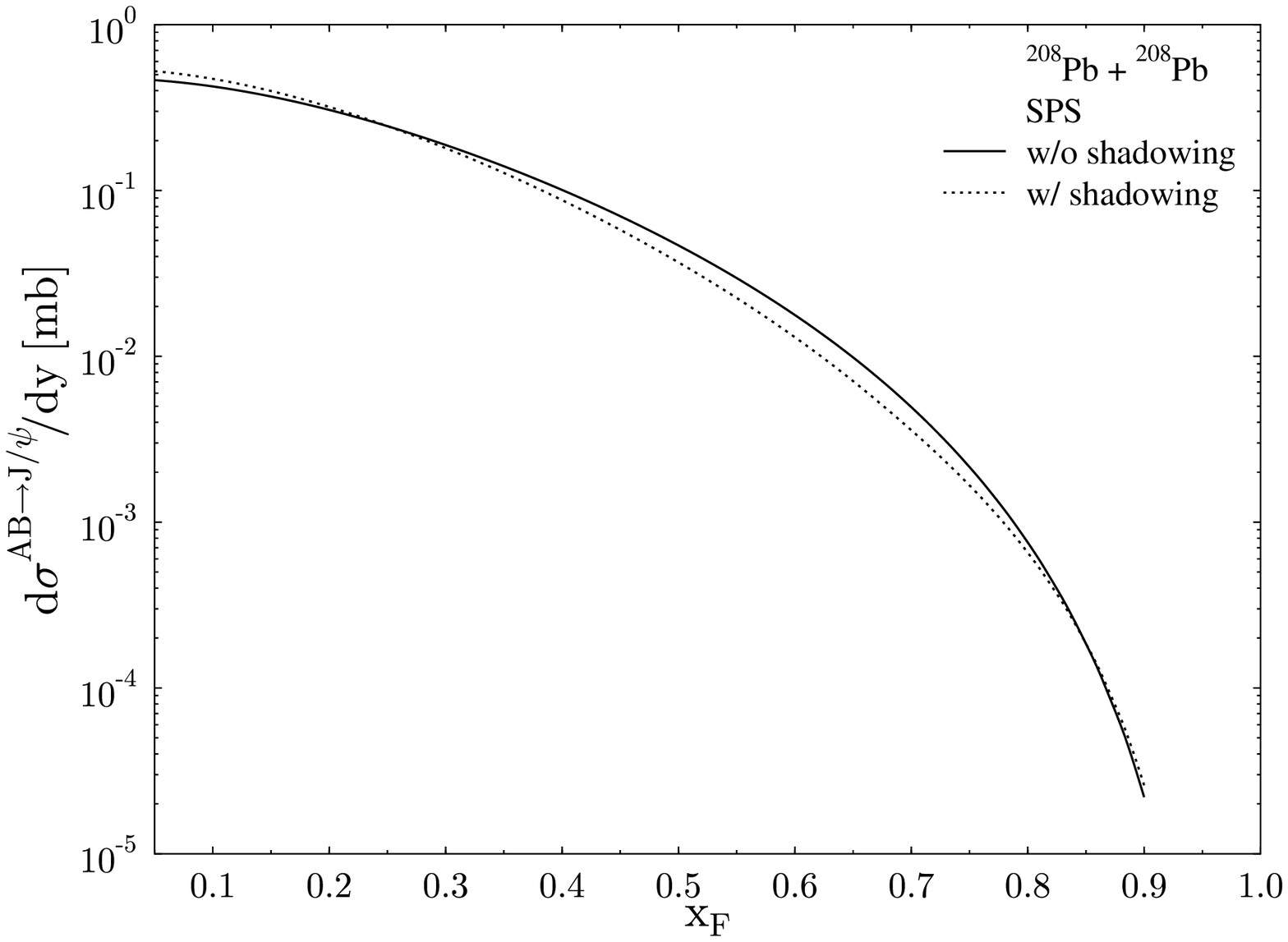,width=10cm}\psfig{figure=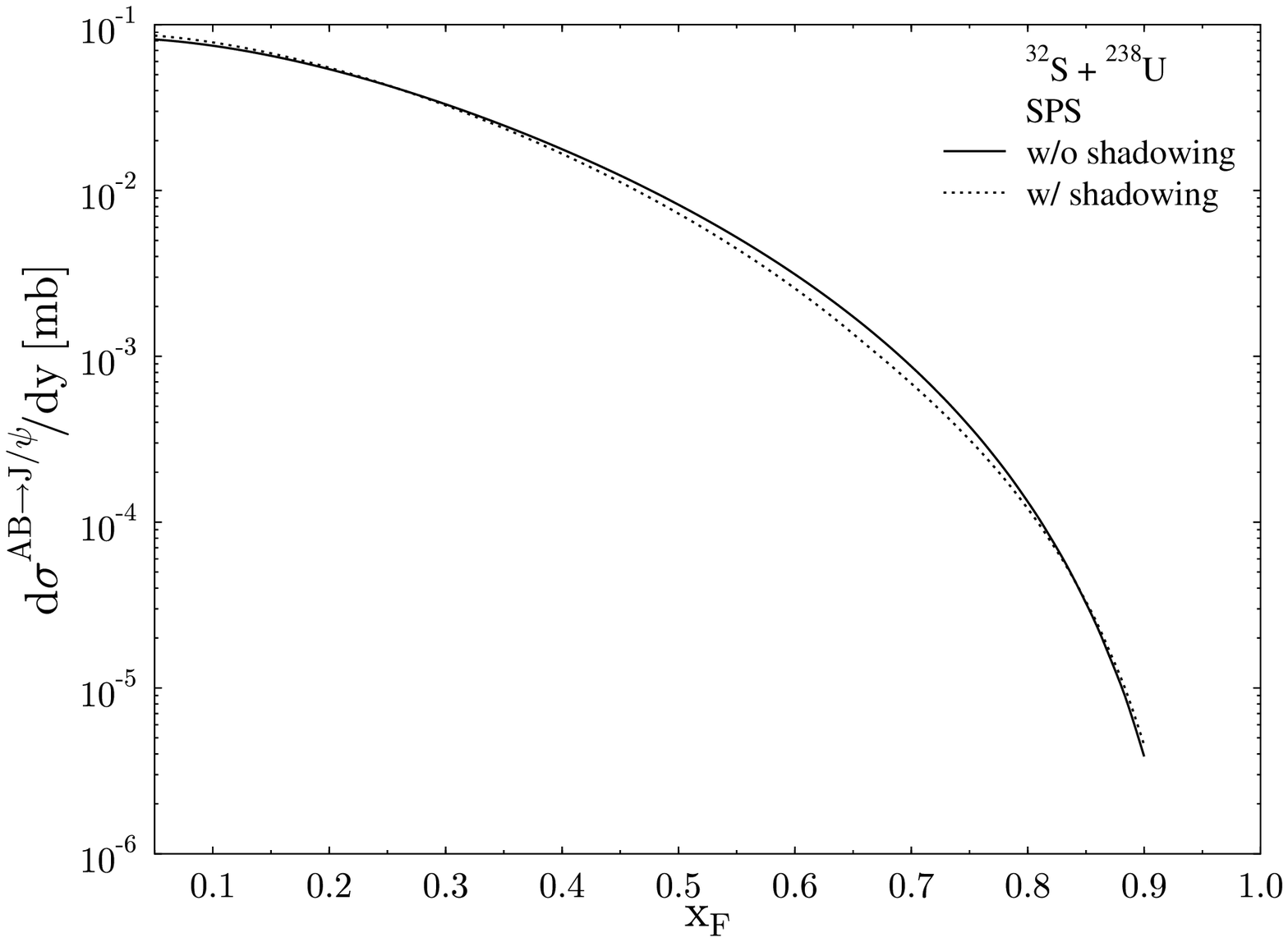,width=10cm}}
\centerline{\psfig{figure=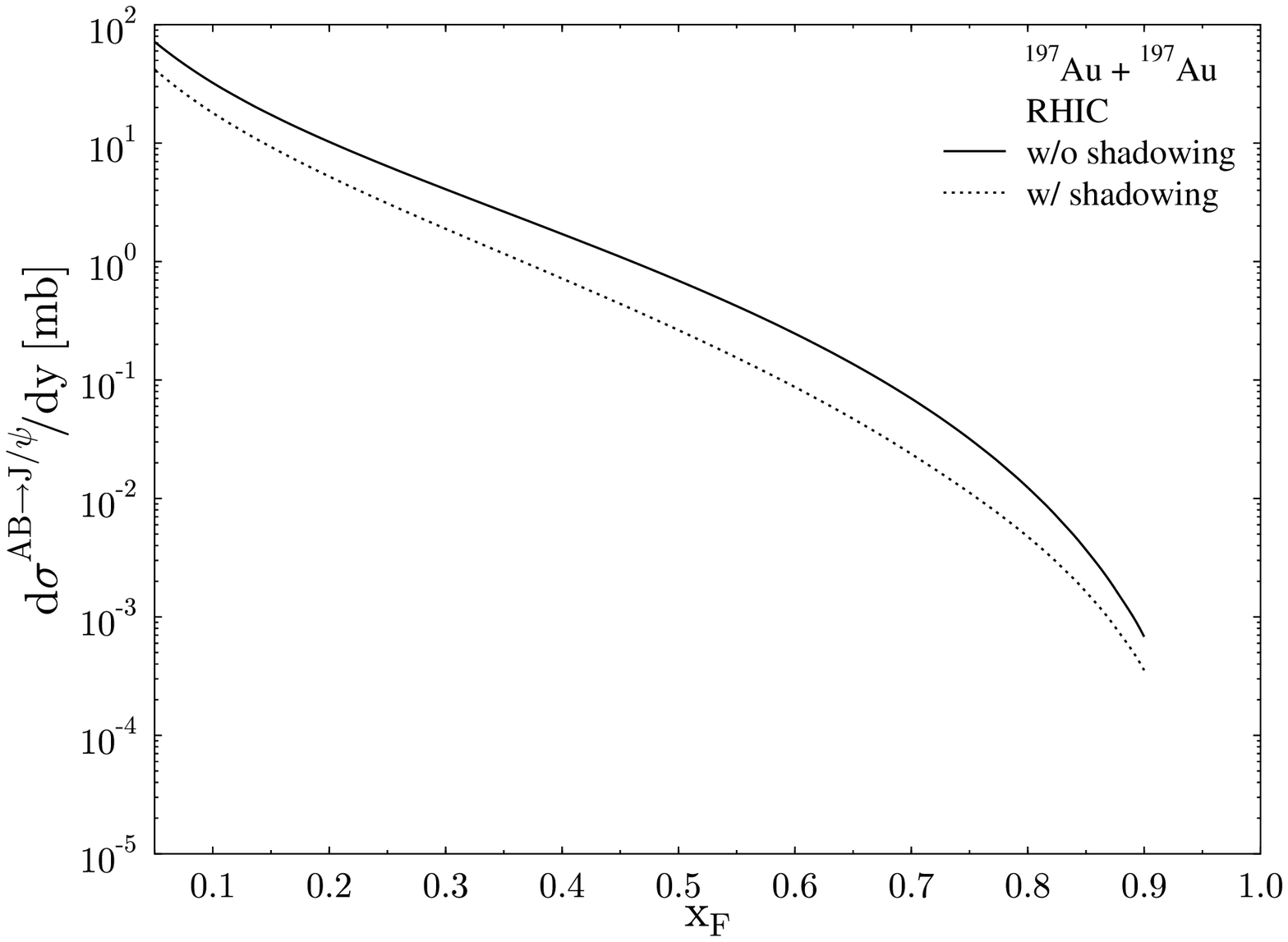,width=10cm}\psfig{figure=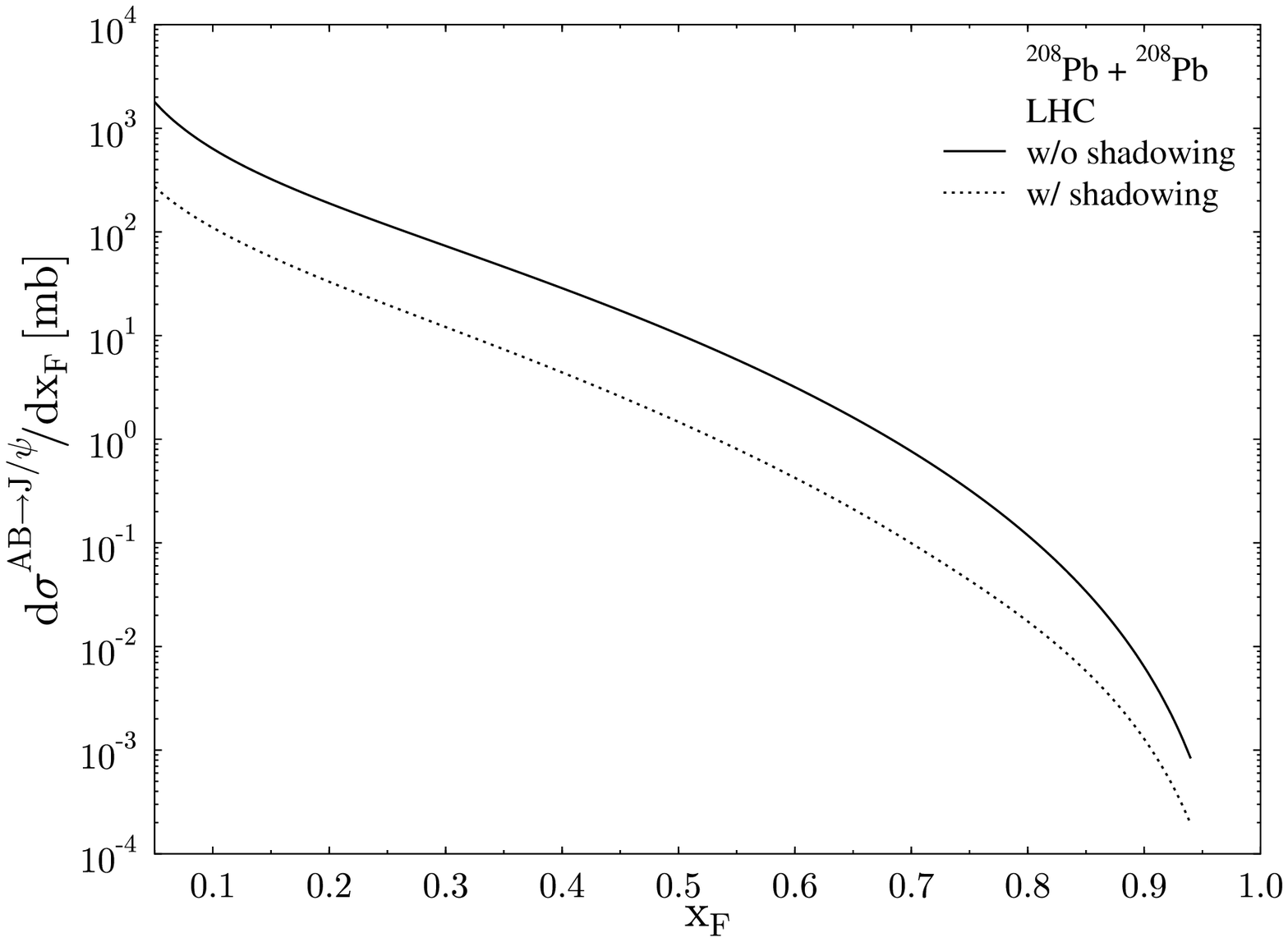,width=10cm}}
\caption{$d\sigma^{AB\rightarrow J/\psi}/dx_F$ for SPS ($Pb+Pb$, $S+U$),
RHIC ($Au+AU$) and LHC ($Pb+Pb$).}        
\label{fig5-181920}
\end{figure}
\begin{figure}
\centerline{\psfig{figure=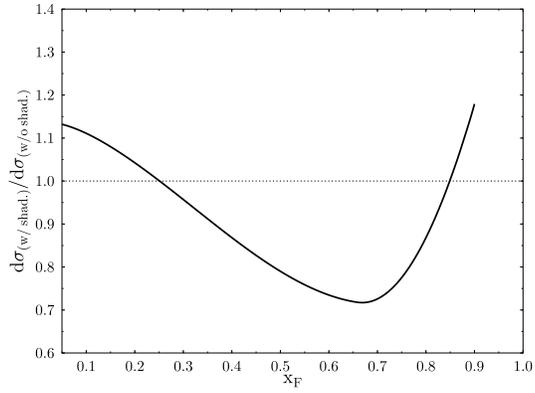,width=10cm}\psfig{figure=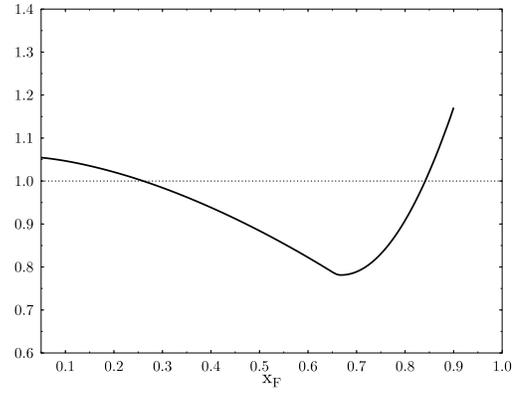,width=10cm}}
\centerline{\psfig{figure=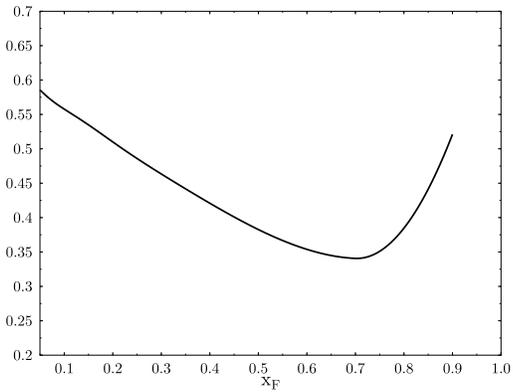,width=10cm}\psfig{figure=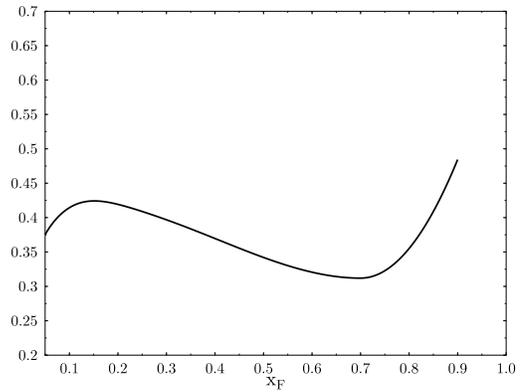,width=10cm}} 
\caption{$\frac{d\sigma^{AB \rightarrow J/\psi}_{w/ ~shad.}}{dx_F}/
\frac{d\sigma^{AB\rightarrow J/\psi}_{w/o ~shad.}}{dx_F}$ 
for SPS ($Pb+Pb$, $S+U$),
RHIC ($Au+AU$) and LHC ($Pb+Pb$).}
\label{fig5-181920ratio}
\end{figure}
\cleardoublepage
\centerline{\bf 5. Conclusions}
From the results shown above one now can draw the following conclusions for the
consequences of the shadowing effects for charmonium production
and suppression at SPS, RHIC and LHC.\\
First, one can conclude from figures \ref{fig5-4} and \ref{fig5-181920ratio} that 
an enhancement of charmonium states produced near midrapidity due to antishadowing at 
$\sqrt s =20$ GeV is predicted (small $x_F$). 
For larger $x_F$,
a clear suppression of the charm cross section to $\approx 70 - 80\%$ of the unshadowed result 
(figure \ref{fig5-181920}) and again a rise at the largest $x_F$ values is predicted
(the latter one due to the Fermi motion effect).\\
For RHIC energies of $\sqrt s =200$ GeV the situation changes; for minijets with 
$1$ GeV $<p_T<6$ GeV 
at midrapidity (or at small $x_F$, respectively) one is completely
in the shadowing region. Here, the shadowed result are reduced by $\approx 45\%$.
At LHC the situation 
is even more dramatic: the ratio of the shadowed cross section to the unshadowed 
cross section at $p_T=1$ GeV is 0.22 which amounts to a suppression of a factor 
$\approx 4.6$.\\ 
Similar effects are observable for $d\sigma ^{AB\rightarrow J/\psi}/dx_F$: 
at small $x_F \approx 0.05$ for RHIC the cross section
is reduced by a factor $d\sigma^{shad.}/d\sigma^{unshad.}\approx 0.58$, and gets suppressed
even more towards larger $x_F$ down to values $\approx 0.35$. 
At LHC one finds a less strong variation over the $x_F$ range with a mean value of
$\approx 0.35$.
In these results one problem is unveiled: the difference
between the (not yet exactly known) gluon ratio $R_G$ and the quark ratio $R_{F_2}$ that, according to the 
calulations in \cite{kumano} increases with increasing mass number. 
If, as it was recently done at CERN-SPS, the future experiments 
at RHIC and LHC compare different combinations of nuclei and derive results
similar to the NA50 data one has to ask 
oneself whether one has detected the plasma or whether the 
detection is that the gluon ratio in {\it not simply} given by $R_{F_2}$, even at small 
$x$. To give clear predictions it is mandatory to control the 
value of $R_G$ at the typical semihard scale $Q_{SH}\approx 2$ GeV with high precision.
Therefore charmonium and bottonium suppression effects can also be due to purely 
geometrical effects, i.e.~shadowing.\\ \\ \\ \\
\centerline{\bf Acknowledgements}
We gratefully appreciated discussions with L.~ Frankfurt and M.~Strikman.


\vskip 0.3 cm
\begin{thebibliography}{99}
%
\bibitem{spieles}  C. Spieles, L. Gerland, N. Hammon, M. Bleicher, S. A. Bass, H. St\"ocker, 
W. Greiner, C. Louren\c{c}o, R. Vogt, {\it Euro. Phys. J.} {\bf A1}, 51 (1998)
\bibitem{nils} N. Hammon, A. Dumitru, H. St\"ocker, W. Greiner, {\it Phys. Rev.} {\bf C57},
3292 (1998)
\bibitem{hwa} R. C. Hwa, J. Pisut, N. Pisutova, nucl-th/9706062
\bibitem{NA50}M. Gonin, NA50 Collaboration, {\it Nucl. Phys.} {\bf A610}, 404c (1996),\\
             C. Louren\c{c}o, NA50 Collaboration, {\it Nucl. Phys.} {\bf A610}, 552c (1996)
\bibitem{matsui} T. Matsui, H. Satz, {\it Phys. Lett.} {\bf B178}, 416 (1986)
\bibitem{neubauer} D. Neubauer, K. Sailer, B. M\"uller, H. St\"ocker, W. Greiner,
{\it Mod. Phys. Lett.} {\bf A4}, 1627 (1989)\\
S. Gavin, M. Gyulassy, A. Jackson, {\it Phys. Lett.} {\bf B207}, 257 (1988)\\
S. Gavin, R. Vogt, {\it Nucl. Phys.} {\bf B345}, 104 (1990)\\
S. Gavin, R. Vogt, {\it Nucl. Phys.} {\bf A610}, 442c (1996)
\bibitem{gavai} R. Gavai, D. Kharzeev, H. Satz, G. Schuler, K. Sridhar, R. Vogt,
hep-ph/9502270
\bibitem{schuler-cern} G.A. Schuler, "Quarkonium production and decays", CERN-preprint,
CERN-TH.7170/94, Feb. '97
\bibitem{lars} L. Gerland, L. Frankfurt, M. Strikman, H. St\"ocker, W. Greiner,
{\it Phys. Rev. Lett.} {\bf 81}, 762 (1998)
\bibitem{EMC} J. J. Aubert, {\it et al.}, {\it Phys. Lett.} {\bf B123}, 275 (1983)
\bibitem{kari} K. J. Eskola, {\it Nucl. Phys.} {\bf B400}, 240 (1993)
\bibitem{glueck} M. Gl\"uck, R. Owens, E. Reya, {\it Phys. Rev.} {\bf D17}, 2324 (1978)
\bibitem{caldwell} D.O. Caldwell, {\it Phys. Rev.} {\bf D7}, 1362 (1973)\\
D.O. Caldwell, {\it Phys. Rev. Lett.} {\bf 42}, 553 (1979)
\bibitem{lonya} L. Frankfurt, M. Strikman, {\it Phys. Rep.} {\bf 160}, 236 (1988)
\bibitem{piller} G. Piller, W. Ratzka, W. Weise, {\it Z. Phys.} {\bf A352}, 427 (1995)
\bibitem{ina} Z. Huang, H. J. Lu, I. Sarcevic, hep-ph/9705250
\bibitem{glr} L. Gribov, E. Levin, M. Ryskin, {\it Phys. Rep.} {\bf 100}, 1 (1983)
\bibitem{mq} A. M\"uller, J. Qiu, {\it Nucl. Phys.} {\bf B268}, 427 (1986)   
\bibitem{close} F.E. Close, J. Qiu, R.G. Roberts, {\it Phys. Rev.} {\bf D40}, 2820 (1989)
\bibitem{E772} D. M. Alde, {\it et al.}, {\it Phys. Rev. Lett.} {\bf 64}, 2479 (1990)
\bibitem{NMC} NM Collaboration, P. Amaudruz, {\it et al.}, {\it Z. Phys.} {\bf C51}, 
387 (1991)
\bibitem{SLAC} R. G. Arnold, {\it et al.}, {\it Phys. Rev. Lett.} {\bf 52}, 727 (1984)
\bibitem{kumano} S. Kumano, K. Umekawa, hep-ph/9803359
\bibitem{lonya1} L. Frankfurt, M. Strikman, S. Liuti, "Fourth Workshop on
Experiments
and Detectors for a Relativistic Heavy Ion Collider", BNL-Report
{\it BNL-52262}, (1990)
\bibitem{mark} M. Strikman, private communication 
\bibitem{emelyanov} V. Emel'yanov, A. Khodinov, S. R. Klein, R. Vogt, hep-ph/9809222
\end{thebibliography}
\end{document}